\documentclass[12pt,paper]{JHEP}

\usepackage{amssymb}
\usepackage{bbm}
\usepackage{amsmath}
\usepackage{amsfonts}
\usepackage{amsthm}

\def\sgn{{\rm sgn}}

\newcommand{\Real}{\textrm{I\!R}}

\newcommand{\e}{\textrm{e}}

\newcommand{\mf}[1]{\mathfrak{#1}}
\newcommand{\mc}[1]{\mathcal{#1}}
\newcommand{\ad}{\textrm{ad}}

\numberwithin{table}{section}

\newtheorem*{lcso}{Lemma on invariant metrics on $CSO$-algebras}

\title{Gauging CSO groups in $N=4$ Supergravity}
\author{Mees de Roo and Dennis B.~Westra\\
     Centre for Theoretical Physics\\
     Nijenborgh 4, 9747 AG Groningen,
     The Netherlands\\
     E-mail: \email{m.de.roo@rug.nl,
     \\ d.b.westra@rug.nl}}
\author{Sudhakar Panda\\
     Harish-Chandra Research Institute \\
     Chatnag Road, Jhusi, Allahabad 211019, India\\
     E-mail: \email{panda@mri.ernet.in}}
\received{\today}               %%
\preprint{\hepth{0606282}\\UG-06/04}

\abstract{We investigate a class of $CSO$-gaugings of $\mc{N}=4$
supergravity coupled to $6$ vector multiplets. Using the
$CSO$-gaugings we do not find a vacuum that is stable against all
scalar perturbations at the point where the matter fields are
turned off. However, at this point we do find a stable
cosmological scaling solution.}

   %check keywords with JHEP

\begin{document}

\section{Introduction\label{Intro}}

Gauged supergravity theories have solutions that may provide a
stringy explanation of cosmological problems. An interesting
example is the fact that gauged $\mc{N}=2$ supergravity has stable
De Sitter solutions
\cite{Fre:2002pd,D'Auria:2002fh,Behrndt:2003bf,Cosemans:2005sj}. A
second class of cosmologically interesting solutions are the
so-called scaling solutions, which might play a role in explaining
the accelerating expansion of the universe (see
\cite{Copeland:1997pi,Liddle:1998pj} and references therein).

In recent work we have investigated the properties of gauged
$\mc{N}=4$ supergravity in four dimensions with the aim of
constructing gaugings which lead to a scalar potential which
allows positive extremum with nonnegative mass matrix
\cite{dRWP1,dRPTW}. No such extrema were found. In the present
paper we extend this work to contracted groups of the CSO type,
and extend the search to include cosmological scaling solutions.

In \cite{dRPTW} we limited ourselves to semisimple gauge groups $G$
with ${\rm dim}\ G \le 12$. Then the field content of the
four-dimensional theory corresponds to that of the $\mc{N}=1,\ d=10$
supergravity, which puts the analysis within a string theory context.
In \cite{dREPW}, we performed a group manifold reduction of the dual
version of $\mc{N}=1,\ d=10$ supergravity, and compared the result with
four-dimensional gauged supergravity. For the group manifold
$SO(3)\times SO(3)$ the resulting gauge group is $CSO(3,0,1)\times
CSO(3,0,1)$. We showed that the effect of this reduction,
including nonzero 3-form fluxes, can also be obtained by directly
gauging the four-dimensional $\mc{N}=4$ theory with the corresponding
$CSO$ group.

In this paper we address $CSO$-gaugings of $\mc{N}=4$ supergravity
with ${\rm dim}\ G \le 12$. $CSO(p,q,r)$ is a contraction of a
special orthogonal group: for $r=0$ they reduce to $SO(p,q)$, if
$r\ne0$ there is an abelian subalgebra of dimension $r(r-1)/2$. To
find consistent $CSO$-gaugings we need to prove a lemma on
invariant metrics on $CSO$-algebras. The main conclusion of the
lemma is that only the $CSO(p,q,r)$-groups with $p+q+r=4$ (we take
$r>0$ in order to have a truly contracted group) give viable
gaugings.

We do not present the most general $CSO$-gauging of $\mc{N}=4$
supergravity. As already mentioned we restrict to $6$ vector
multiplets. In reference \cite{Schon:2006kz} the most general
gaugings of $\mc{N}=4$ supergravity are discussed and
characterized by a set of parameters $\left\{ \xi_{\alpha M},
f_{\alpha KLM}; \; \alpha=1,2\,;\,1\leq K,L,M\leq 12 \right\}$
that need to satisfy a set of constraints. Our gaugings correspond
to the subset of gaugings where $\xi_{\alpha M}=0$.

The paper is organized as follows. In section \ref{scalars} we
discuss the scalar fields of $\mc{N}=4$ matter-coupled
supergravity. In section \ref{CSO} we discuss the gauging of
$\mc{N}=4$ supergravity coupled to $6$ vector multiplets; we
briefly review the concept of $CSO$-groups, or actually their Lie
algebras $\mf{cso}(p,q,r)$, present the lemma on invariant metrics
on $\mf{cso}$-algebras and discuss the $SU(1,1)$-angles. In
section \ref{Pot} we review some results from \cite{dRPTW} and in
section \ref{analysis} we apply this to the $CSO$-gaugings. As in
the case of semisimple groups we do not obtain a positive extremum
with nonnegative mass matrix. In section \ref{scaling} we show
that a cosmological scaling solution exists in $\mc{N}=4$ $CSO$
gauged supergravity.

\section{The scalars of $\mc{N}=4$ supergravity}\label{scalars}

We consider gauged $\mc{N}=4$ supergravity coupled to $n$ vector
multiplets \cite{MdRPW1}. The scalars parameterize an
$SO(6,n)/SO(6)\times SO(n) \times SU(1,1)/U(1)$ coset and can be
split in the $6n$ scalars of the matter multiplets, which
parameterize $SO(6,n)/SO(6)\times SO(n)$-coset, and the two
scalars of the supergravity multiplet, which parameterize an
$SU(1,1)/U(1)$-coset.

The $SU(1,1)$-scalars from the supergravity multiplet are denoted
$\phi_\alpha$, $\alpha=1,2$, and take complex values. When we
define $\phi^1=(\phi_1)^*$ and $\phi^2 = -(\phi_2)^*$, the
constraint that restrict them to the $SU(1,1)/U(1)$-coset reads
\begin{equation}
\label{conphi}
   \phi^\alpha\phi_\alpha = |\phi_1|^2 - |\phi_2|^2 = 1\,.
\end{equation}
A convenient parametrization of the $SU(1,1)$-scalars is obtained
by using the $U(1)$-symmetry to take $\phi_1$ real:
\begin{equation}
\phi_1 = \frac{1}{\sqrt{1-r^2}}\,, \ \ \ \phi_2 =
\frac{r\e^{i\varphi}}{\sqrt{1-r^2}}\,.\label{paramSU}
\end{equation}
The kinetic term of the $SU(1,1)$-scalars then becomes
\begin{equation}
\mc{L}_{kin}(r,\varphi)= - \frac{1}{(1-r^2)^2} \left(
\partial_\mu r\partial^{\,\mu} r + r^2\partial_\mu \varphi \partial^{\,\mu}
\varphi \right) \, .
\end{equation}

The $SO(6,n)$-scalars from the matter multiplets are denoted
$Z_{a}{}^R$, $a=1,\ldots,6$ and $R=1,\ldots,6+n$, and they take
real values. The constraint that restricts the $SO(6,n)$-scalars
to the $SO(6,n)/SO(6)\times SO(n)$-coset is
\begin{equation}
Z_a{}^R\eta_{RS} Z_b{}^S = -\delta_{ab}\,,
\end{equation}
where $\eta_{RS}$ are the components of the invariant metric in
the vector representation of $SO(6,n)$ in a basis such that
\begin{equation}
 \eta = \textrm{diag}(-1,\ldots,-1,+1,\ldots,+1),
\label{eta}
\end{equation}
with six negative
entries and $n$ positive entries. Hence the scalars $Z_a{}^R$ can
be viewed as the upper six rows of $SO(6,n)$-matrix. We define
$Z^{RS}=Z_a{}^{R}Z_a{}^S$ and note that
$\delta_{R}{}^{S}+2Z_{R}{}^S$ is an $SO(6,n)$-matrix, where the
indices are raised and lowered with the metric $\eta_{RS}$.

In this paper we restrict ourselves to $n=6$, which makes contact
with string theory. From \cite{MS} we find a convenient
parametrization of the coset $SO(6,6)/SO(6)\times SO(6)$; we write
$Z_a{}^R = (X,Y)_a{}^R$, where $X$ and $Y$ are $6\times
6$-matrices and put
\begin{equation}
\begin{split}
X &=\tfrac{1}{2} \left( G + G^{-1} + BG^{-1} - G^{-1}B -
BG^{-1}B\right) \, ,\\
Y&= \tfrac{1}{2}\left( G - G^{-1} - BG^{-1} - G^{-1}B -
BG^{-1}B\right) \, ,
\end{split}
\end{equation}
where $G$ is an invertible symmetric $6\times 6$-matrix and $B$ is
an antisymmetric $6\times 6$-matrix. It is convenient to split the
indices $R,S,\ldots$ of $\eta_{RS}$ in $A,B,\ldots=1,\ldots,6$,
($\eta_{AB}=-\delta_{AB}$) and $I,J,\ldots=7,\ldots,12$,
($\eta_{IJ}=+\delta_{IJ}$). Hence $Z_a{}^A = X_a{}^A$ and
$Z_a{}^I=Y_a{}^{I-6}$. We define a $6\times 6$-matrix containing
the independent degrees of freedom of the $SO(6,6)$-scalars by
$P=G+B$ and denote its components by $P_{ab}$, where $1\leq
a,b,\leq 6$. The kinetic term of the independent scalars $P_{ab}$
then reads:
\begin{equation}
\mc{L}_{kin}(P_{ab}) = -\tfrac{1}{2} \partial_\mu P_{ab}
\partial^{\,\mu} P_{ab} \,.
\end{equation}

There is a certain freedom in coupling the vector multiplets: for
each multiplet, labelled by $R$, we can introduce an
$SU(1,1)$-element, of which only a single angle $\alpha_R$ turns
out to be important. These angles $\alpha_R$ can be reinterpreted
as a modification of the $SU(1,1)$-scalars coupling to the
multiplet $R$ in the form
\begin{equation}
    \phi^1_{(R)} = e^{i\alpha_R}\phi^1\,,\quad
    \phi^2_{(R)} = e^{-i\alpha_R}\phi^2\,,\quad
    \Phi_{(R)} = e^{i\alpha_R}\phi^1 + e^{-i\alpha_R}\phi^2\,.
\end{equation}
The kinetic term of the vector fields $A_{\mu}^{R}$ is
\begin{equation}
\mc{L}_{kin}(A_{\mu}^{R}) =
-\frac{\eta_{RS}+2Z_{RS}}{4|\Phi_{(R)}|^2}F_{\mu\nu}^{R}F^{S\mu\nu}
\, ,
\end{equation}
where $F_{\mu\nu}^{R}$ is the nonabelian field strength of
$A_{\mu}^R$.

In this paper we are mainly interested in a special point of the
$SO(6,6)/SO(6)\times SO(6)$-manifold, namely the point where the
matter multiplets are `turned off'. This point is denoted $Z_0$
and corresponds to the identity point of the coset
$SO(6,6)/SO(6)\times SO(6)$, that is, $Z_0 \cong SO(6)\times
SO(6)$. In our parametrization we have at $Z_0$: $ X=\mathbbm{1}$,
$Y=0$ and $P_{ab}=\delta_{ab}$.

\section{CSO gaugings\label{CSO}}

In the context of maximal supergravities $CSO$-groups have been
used to construct gauged supergravities, see e.g.
\cite{hull1,hull2,hull3,an1,an3}. By truncating the
four-dimensional $\mc{N}=8$ theory to an $\mc{N}=4$ theory one
obtains four-dimensional $\mc{N}=4$ supergravities with a
$CSO$-gauging\cite{hull1,hull3}. The definition of $CSO$-algebras
as outlined below is similar to the discussion in \cite{an1}.

Let $\mf{g}$ be a real Lie algebra, then $\mf{g}$ is admissible as
a gauge algebra of $\mc{N}=4$ supergravity if and only if there
exists a basis of generators $T_R$ of $\mf{g}$ such that the
structure constants defined by $[T_R,T_S]=f_{RS}{}^U T_U$ satisfy
\begin{equation}
f_{RS}{}^T\eta_{TU} +
f_{RU}{}^T\eta_{TS}=0\,,\label{etaconstraint}
\end{equation}
with $\eta$ as defined in (\ref{eta}).
We define a symmetric nondegenerate bilinear form $\Omega$ on the
Lie algebra $\mf{g}$ by its action on the basis elements $T_R$
through
\begin{equation}
\Omega(T_R,T_S) = \eta_{RS}\,.
\end{equation}
The constraint (\ref{etaconstraint}) then is equivalent to
demanding that the form $\Omega$ is invariant under the adjoint
action of the Lie algebra $\mf{g}$ on itself. From now on we write
`metric' for `nondegenerate bilinear symmetric form'.

On complex simple Lie algebras there exists only a one-parameter
family of invariant metrics and every invariant metric is
proportional to the Cartan--Killing metric. For simple real
algebras of which the complex extensions is simple there exists up
to multiplicative factor only one invariant metric, given by the
Cartan--Killing metric. However, for simple real Lie algebras of
which the complex extension is not simple, there exists a
two-parameter family of invariant metrics. This can be seen from
the fact that if the complex extension is not simple, it is of the
form $\mf{m}\oplus\mf{m}$, with $\mf{m}$ a complex simple Lie
algebra.

For Lie algebras of the type $\mf{cso}(p,q,r)$ (for definitions,
see section \ref{csodef}) the situation is more delicate. The
criterion of nondegeneracy turns out to be very restrictive. We
have the following useful lemma:

\begin{lcso}
The Lie algebra $\mf{cso}(p,q,r)$ with $r>0$ admits an invariant
nondegenerate symmetric bilinear form (i.e. an invariant metric)
only if
\begin{equation}
  {\rm (1)}\ p+q +r =2\ {\rm or}\  {\rm (2)}\  p+q+r=4\,.
\label{lemma}
\end{equation}
\end{lcso}

\noindent Since the algebras $\mf{cso}(1,0,1)\cong
\mf{cso}(0,0,2)\cong\mf{cso}(0,0,4)\cong \mf{u}(1)$ are abelian,
the structure constants are zero and give therefore rise to
trivial gaugings. Hence, we focus on the CSO-algebras of the type
$\mf{cso}(p,q,r)$ with $p+q+r=4$ and $0<r<4$.

\subsection{Lie algebras of the type
$\mf{cso}(p,q,r)$}\label{csodef}

In the vector representation the Lie algebra $\mf{so}(p,q+r)$
admits a set of basis elements $J_{AB}=-J_{BA}$, $1\leq A,B \leq
p+q+r$ satisfying the commutation relation:
\begin{equation}
[J_{AB}, J_{CD}] = g_{BC}J_{AD} + g_{AD}J_{BC} - g_{AC}J_{BD} -
g_{BD}J_{AC}\, ,
\end{equation}
where $g_{AB}$ are the entries of the diagonal matrix with $p$
eigenvalues $+1$ and $q+r$ eigenvalues $-1$.

We split the indices\footnote{The splitting of indices in this
case is not related to the splitting of the indices of the
$SO(6,6)$-scalars introduced in section \ref{scalars}.}
$A,B,\ldots$ into indices $I,J,\ldots$ running
from $1$ to $p+q$ and indices $a,b,\ldots$ running from $p+q+1$ to
$p+q+r$. The Lie algebra $\mf{so}(p,q+r)$ splits as a vector space
direct sum $\mf{so}(p,q+r) = \mf{so}(p,q)\oplus \mc{V}\oplus
\mc{Z}$, where the elements $J_{IJ}$ span the $\mf{so}(p,q)$
subalgebra, the elements $J_{Ia}=-J_{aI}$ span the subspace
$\mc{V}$ and the elements $J_{ab}$ span the subalgebra $\mc{Z}$.
The subspace $\mc{V}$ consists of $r$ copies of the vector
representation of the subalgebra $\mf{so}(p,q)$, whereas the
subalgebra $\mc{Z}$ consists of singlet representations of
$\mf{so}(p,q)$. The commutation relations are schematically given
by:
\begin{equation}
\begin{split}
&\left[ \mf{so}(p,q)\, ,\mc{V} \right] \subset \mc{V}\, , \quad
\quad \quad \quad \quad \ \
\left[\mc{V}\, ,\mc{V} \right] \subset \mc{Z}\oplus\mf{so}(p,q)\, , \\
&\left[ \mf{so}(p,q)\, ,\mc{Z} \right] \subset 0\, ,  \quad \quad
\ \  \quad \quad \quad \left[
\mc{Z}\, ,\mc{V} \right] \subset \mc{V}\, , \\
&\left[ \mf{so}(p,q)\, ,\mf{so}(p,q) \right] \subset
\mf{so}(p,q)\, , \ \ \left[ \mc{Z}\, ,\mc{Z} \right] \subset
\mc{Z}\, .
\end{split}
\end{equation}

We define for any $\xi\in\Real$ a linear map
$T_\xi:\mf{so}(p,q+r)\rightarrow \mf{so}(p,q+r)$ by its action on
the subspaces:
\begin{equation}
\begin{split}
 x\in \mf{so}(p,q)\, , \quad &T_\xi : x\mapsto x \,,\\
 x\in \mc{V}\, , \quad\quad &T_\xi: x\mapsto \xi x \,,\\
 x\in \mc{Z}\, , \quad\quad &T_\xi: x\mapsto \xi^2 x\, .
\end{split}
\end{equation}
If $\xi\neq 0,\infty$ the map $T_\xi$ is a bijection. The maps
$T_0$ and $T_\infty$ give rise to so-called contracted Lie
algebras.

We define the limits $T_0 (\mf{so}(p,q)) =
\mf{s}\cong\mf{so}(p,q)$, $T_0(\mc{V})=\mf{r}$ and
$T_0(\mc{Z})=\mf{z}$. The Lie algebra $\mf{cso}(p,q,r)$ is defined
as $T_0(\mf{so}(p,q+r))$. Hence we have
$\mf{cso}(p,q,r)=\mf{so}(p,q)\oplus\mf{r}\oplus\mf{z}$ and the
commutation rules are of the form
\begin{equation}
 \left[ \mf{s}\, ,\mf{s} \right] \subset \mf{s}\, , \  \quad
\left[ \mf{r} \, ,\mf{r} \right] \subset\mf{z} \, , \  \quad
\left[ \mf{s}\, , \mf{r}\right] \subset \mf{r}\, , \  \quad
\left[\mf{r} \, , \mf{z}\right]= \left[\mf{s} \, , \mf{z}\right] =
\left[\mf{z} \, ,\mf{z} \right] = 0\, .
\end{equation}

We mention some special cases and properties. If $r=0$ the
construction is trivial and therefore we take $r>0$. If $p+q=1$ we
have $\mf{s}=0$ and if $p+q=r=1$ also $\mf{z}=0$ and we have
$\mf{cso}(1,0,1)\cong \mf{cso}(0,1,1)\cong\mf{u}(1)$. If $p+q = 2$
the Lie algebra $\mf{s}$ is abelian and if $p+q>2$ the Lie algebra
$\mf{s}$ is semisimple and the vector representation is
irreducible. Hence if $p+q>2$ we have $[\mf{s},\mf{r}]\cong
\mf{r}$. If $r=1$ we have $\mf{z}=0$ and the Lie algebra is an
In\"on\"u--Wigner contraction.

From the construction follows a convenient set of basis elements
of $\mf{cso}(p,q,r)$. The elements $S_{IJ} = -S_{JI}$ are the
basis elements of the subalgebra $\mf{s}$, the elements $v_{Ia}$
are the basis elements of $\mf{r}$ and the elements
$z_{ab}=-z_{ba}$ are the basis elements of $\mf{z}$. The only
nonzero commutation relations are:
\begin{equation}
\begin{split}
\left[ S_{IJ}, S_{KL} \right] &= \tilde g_{JK}S_{IL} -
        \tilde g_{IK}S_{JL}-\tilde g_{JL}S_{IK}+\tilde g_{IL}S_{JK}\,, \\
        \left[ S_{IJ}, v_{Ka} \right] &=
        \tilde g_{JK}v_{Ia}-\tilde g_{IK}v_{Ja}\,,\\
        \left[ v_{Ia},v_{Jb} \right] &= \tilde g_{IJ}Z_{ab}\, .
        \label{altdefcso}
\end{split}
\end{equation}
The numbers $\tilde g_{IJ}$ are the elements of the diagonal
metric with $p$ eigenvalues $+1$ and $q$ eigenvalues $-1$. The
commutation relations (\ref{altdefcso}) can also be taken as the
definition of the Lie algebra $\mf{cso}(p,q,r)$.

\subsection{Choosing the $SU(1,1)$-angles}

In general the gauge algebra $\mf{g}$ can be decomposed as a
direct sum $\mf{g}=\mf{g}_1\oplus\mf{g}_2\oplus\ldots$ and it is
clear that the $SU(1,1)$-angles can be different on different
factors $\mf{g}_i$. With each generator $T_R$ of the gauge algebra
$\mf{g}$ we associate a gauge field $A_{\mu}^{R}$ and an
$SU(1,1)$-angle $\alpha_R$. The gauge group rotates the gauge
fields associated to the same factor into each other. All the
generators that can be obtained by rotating the generator $T_R$
need to have the same $SU(1,1)$-angle $\alpha_R$ for the gauge
group to be a symmetry. Hence along the gauge orbit of $T_R$,
denoted by $\Gamma[ T_R ]$ and defined by
\begin{equation}
\Gamma\left[ T_R \right] = \left\{ \e^{\ad A} (T_R) | A\in\mf{g}
\right\}\, ,\label{gauge-orbit}
\end{equation}
the $SU(1,1)$-angle has to be constant. If
$\Gamma[T_R]\cap\Gamma[T_S]\neq 0$ we need $\alpha_R=\alpha_S$.
For semisimple groups the gauge orbits are the simple factors and
hence with each simple factor we associate a single
$SU(1,1)$-element.

For the algebras $\mf{cso}(2,0,2)$, $\mf{cso}(1,1,2)$ the gauge
orbit of $\mf{s}$, which is one-dimensional, is
$\mf{s}\oplus\mf{r}$ and the gauge-orbit of $\mf{r}$ is
$\mf{r}\oplus\mf{z}$. For the algebras $\mf{cso}(3,0,1)$ and
$\mf{cso}(2,1,1)$ the gauge orbit of every element of $\mf{s}$ is
the whole Lie algebra. Finally, for $\mf{cso}(1,0,3)$ the gauge
orbit of each element $\mf{r}$ is contained in
$\mf{r}\oplus\mf{z}$ and all gauge orbits overlap. Hence for all
$CSO$-type algebras under consideration the $SU(1,1)$-angles have
to be constant over the whole Lie algebra $\mf{cso}(p,q,r)$.

\subsection{The embedding of CSO-algebras in $SO(6,6)$}

The $CSO$-algebras that are admissible are the $\mf{cso}(p,q,r)$
with $p+q+r=1$, and since for $r=0$ the algebra is semisimple, we
only consider $r>0$.

To find a basis such that (\ref{etaconstraint}) is satisfied on
the structure constants we first construct any basis for the Lie
algebra and find the invariant metric $\Omega$, which in most
cases can be cast in a simple form. The second step is to find a
basis-transformation such that in the new basis $\Omega$ is
diagonalized with all eigenvalues $\pm 1$. Then the structure
constants are calculated in this basis, and by construction they
satisfy (\ref{etaconstraint}). This procedure is not unique and it
is easy to see that any $SO(6,6)$-transformation on the structure
constants leaves the constraint (\ref{etaconstraint}) invariant.
However, we are not trying to be completely exhaustive. On the
other hand, an $SO(6,6)$-transformation can be seen as a rotation
on the scalar fields $Z_a{}^R$ and has the physical interpretation
of turning on the matter fields (if the rotation is not contained
in the subgroup $SO(6)\times SO(6)$).

For all $CSO$-algebras under consideration the dimension is six
and the invariant metric has signature $+++---$ (see section
\ref{proof}). This implies that precisely two $CSO$-algebras can
be embedded into the vector representation of $SO(6,6)$.

There is a $\mathbbm{Z}_2$-freedom in choosing the embedding into
$SO(6,6)$: for a given invariant metric $\Omega$ the eigenvectors
with positive eigenvalues can be embedded either in the subspace
spanned by the generators $T_A$ where $\eta_{AB} = - \delta_{AB}$
or in the subspace spanned by the generators $T_I$ where
$\eta_{IJ}=+\delta_{IJ}$. This difference in embedding can result
in a physical difference that modifies the potential. To
distinguish between the two kinds of structure constants resulting
from the difference in embedding we denote one embedding as
$CSO(p,q,r)_+$ and the other as $CSO(p,q,r)_-$. In contrast to the
case of semisimple gaugings (where one can use the Cartan--Killing
metric to choose a sign-convention), the procedure of assigning a
plus or minus to the gauging is arbitrary, since if $\Omega$ is an
invariant metric, then also $-\Omega$ is an invariant metric that
interchanges the plus- and minus-type of gauging. In appendix
\ref{structconst} we present the structure constants for the
different embeddings. We note that for the Lie algebras
$\mf{cso}(2,0,2)$ and $\mf{cso}(1,1,2)$ the structure constants of
the plus-embedding and minus-embedding are the same, hence no
distinction will be made for these algebras.

Having obtained a set of structure constants that satisfies the
constraint (\ref{etaconstraint}) we return to the $\mc{N}=4$
supergravity and use the structure constants as input to
investigate the potential. In section \ref{Pot} we present the
details of the potential that are used in the analysis and in
section \ref{analysis} we present the analysis of the potential
with the $CSO$-gaugings. To finish this section, we give a proof
of the lemma on invariant metric on $CSO$-algebras.

\subsection{Proof of the lemma}\label{proof}

The proof consists of two parts. In the first part
({\bf Part I} below) we prove for
all but the $CSO$-algebras listed in (\ref{lemma}) that no invariant
metric exists. We do this by assuming a bilinear form $\Omega$ is
invariant and then prove it is degenerate. In the second part
({\bf Part II}) we
give  the invariant
metrics for the $CSO$-algebras listed in (\ref{lemma}).

The first part uses the concepts of isotropic subspaces and
Witt-indices. For a bilinear symmetric form $B$ on a real vector
space $V$, an isotropic subspace is a subspace $W$ of $V$ on which
$B$ vanishes. The maximal isotropic subspace is an isotropic
subspace with the maximal dimension. The dimension of the maximal
isotropic subspace is the Witt-index of the pair $(B,V)$ and is
denoted $m_W$.

If $B$ is nondegenerate and the dimension of $V$ is $n$, one can
always choose a basis in which $B$ has the matrix form
\begin{equation}
B=\begin{pmatrix} \mathbbm{1}_{p\times p} & 0 & \\ 0 & 0 &
\mathbbm{1}_{r\times r} \\ 0 & \mathbbm{1}_{r \times r} & 0 \\
\end{pmatrix}\, ,  \ \ \textrm{for }p,r \textrm{ with } p+2r = n\, .
\end{equation}
This clearly shows that the Witt-index is $r$. Hence we have the
inequality: $m_W \leq [n/2]$.

If the center $\mf{z}$ is nonzero we have
$[\mf{r},\mf{r}]=\mf{z}$, that is, for every $z \in \mf{z}$ there
are $v_i,w_i\in\mf{r}$ such that $\sum_i[v_i,w_i]=z$. Hence if
$z,z'$, with $z=\sum_i[v_i,w_i]$ and $v_i,w_i\in\mf{r}$, we have
$\Omega(z,z') = \sum_i\Omega([v_i,w_i],z') = \sum_i
\Omega(v_i,[w_i,z']) = 0$ and hence the center $\mf{z}$ is
contained in the maximal isotropic subspace. Hence if the
dimension of $\mf{z}$ exceeds half the dimension of the Lie
algebra, any invariant symmetric bilinear form is necessarily
degenerate.
\begin{center}
\subsection*{Part I}
\end{center}
We split part I in six different cases. For every case we assume
an invariant symmetric bilinear form $\Omega$ exists and prove
degeneracy. We use the same decomposition as in section
\ref{csodef}, $\mf{g}=\mf{s}\oplus\mf{r}\oplus\mf{z}$, with
$\mf{g}$ a $CSO$-type Lie algebra, and the standard commutation
relations (\ref{altdefcso}).

\begin{center}
\underline{\textbf{$\mf{cso}(p,q,r)$ with $p+q>2$ and $r>1$}}
\end{center}

We have $[\mf{s},\mf{s}]=\mf{s}$, $[\mf{r},\mf{r}]=\mf{z}$ and
$[\mf{s},\mf{r}]=\mf{r}$. We prove that $\mf{z}$ is perpendicular
to the whole algebra with respect to $\Omega$, which implies that
$\Omega$ is degenerate.

The center $\mf{z}$ is perpendicular to itself since it is nonzero
and thus defines an isotropic subspace. For every $v\in \mf{r}$
there are $j_i\in\mf{s}$ and $w_i\in\mf{r}$ such that
$\sum_i[j_i,w_i]=v$. Hence for such $v$ and $z\in\mf{z}$ we have
$\Omega(v,z)=\sum_i \Omega(j_i,[w_i,z])=0$ and $\Omega$ is zero on
$\mf{z}\times\mf{r}$. Since $\mf{s}$ is semisimple a similar
argument shows that $\Omega$ is zero on $\mf{z}\times \mf{s}$ and
then $\mf{z}$ is orthogonal to the whole Lie algebra with respect
to $\Omega$.

\begin{center}
\underline{\textbf{$\mf{cso}(p,q,r)$ with $p+q=1$ and $r>3$}}
\end{center}

We have $\mf{s}=0$ and $\dim \mf{r}=r$ and $\dim \mf{s}=r(r-1)/2$.
The dimension of the center becomes too large for $\Omega$ to be
nondegenerate if $r(r-1)/ 2 > r(r+1)/4 $. It follows that if $r>3$
there is no invariant metric.

\begin{center}
\underline{\textbf{$\mf{cso}(p,q,r)$ with $p+q=1$ and $r=2$}}
\end{center}

From the commutation relations (\ref{altdefcso}) we see that we
can choose a basis $e,f,z$ such that the only nonzero commutator
is $[e,f]=z$. We have $\Omega(z,z)=0$, but also $\Omega(e,z) =
\Omega(e,[e,f]) = \Omega([e,e],f)= 0$. Similarly $\Omega(z,f)=0$
and thus $z$ is perpendicular to the whole algebra and $\Omega$ is
degenerate.

\begin{center}
\underline{\textbf{$\mf{cso}(p,q,r)$ with $p+q=2$ and $r=1$}}
\end{center}

The Lie algebras $\mf{cso}(1,1,1)$ and $\mf{cso}(2,0,1)$ have zero
center and hence $[\mf{r},\mf{r}]=0$. For every $x\in\mf{r}$ there
are $y_i\in \mf{r}$ and $A_i\in\mf{s}$ such that
$x=\sum_i[A_i,y_i]$. Therefore we have for such $x,y_i,A_i$ and $v
\in\mf{r}$ : $\Omega( x,v)  = \sum_i\Omega( A_i, [y_i,v]) = 0$.
Thus $\mf{r}$ is an isotropic subspace of dimension 2, whereas the
dimension of the Lie algebra is 3.

\begin{center}
\underline{\textbf{$\mf{cso}(p,q,r)$ with $p+q=2$ and $r>2$}}
\end{center}

We choose a basis $\left\{ j, e_a, f_a, z_{ab}\right\}$, where
$j\in\mf{s}$, $e_a,f_a\in\mf{r}$ and $z_{ab}=-z_{ba}\in \mf{z}$
and $1\leq a,b\leq r$. In terms of the basis elements in
(\ref{altdefcso}) we have $j=J_{12}$, $e_a = v_{1a}$, $f=v_{2a}$.
The only nonzero commutation relations are
\begin{equation}
 \left[ j,e_a \right] = f_a\, , \ \ \left[ j,f_a\right] = \sigma
e_a\, , \ \ \left[ f_a,f_b \right] = \sigma z_{ab}\, \ \  \left[
e_a,e_b \right] = z_{ab}\, , \label{cso11Ncomm}
\end{equation}
where $\sigma=+1$ for $\mf{cso}(1,1,r)$ and $\sigma=-1$ for
$\mf{cso}(2,0,r)$.

From the commutation relations (\ref{cso11Ncomm}) one deduces that
the subspace spanned by the elements $e_a$ and $z_{ab}$ defines an
isotropic subspace of dimension $r(r+1)/2$. The dimension of this
isotropic subspace exceeds half the dimension of the Lie algebra
if $r>2$.

\begin{center}
\underline{\textbf{$\mf{cso}(p,q,r)$ with $p+q>3$ and $r=1$}}
\end{center}

The Lie algebras in this class have zero center and hence
$[\mf{r},\mf{r}]=0$. We have $\mf{r}=[\mf{s},\mf{r}]$,
$\mf{s}=[\mf{s},\mf{s}]$ and $\mf{s}$ is semisimple. It follows
that $\Omega$ is zero on $\mf{r}\times\mf{r}$ and $\Omega$
coincides with the Cartan--Killing metric of $\mf{s}$ on
$\mf{s}\times\mf{s}$. Hence we are interested in $\Omega$ on
$\mf{r}\times\mf{s}$.

From (\ref{altdefcso}) we see that we can choose a basis $\left\{
S_{IJ}, v_{I} \right\}$, where $1\leq I,J \leq p+q$, and the only
nonzero commutation relations are:
\begin{equation}
\begin{split}
\left[ S_{IJ}, S_{KL} \right] &= \eta_{JK}S_{IL} -
        \eta_{IK}S_{JL}-\eta_{JL}S_{IK}+\eta_{IL}S_{JK} \\
        \left[ S_{IJ}, v_{K} \right] &=
        \eta_{JK}v_{I}-\delta_{IK}v_{J}\, .
\end{split}
\end{equation}

We define $\Omega_{IJK}=\Omega(v_I,S_{JK})=-\Omega_{IKJ}$.
Invariance requires $\Omega([S_{IJ}, v_{K}], S_{LM}) = -
\Omega(v_{K}, [S_{IJ},S_{LM}])$, from which we obtain:
\begin{equation}
\eta_{JK}\Omega_{ILM} - \eta_{IK}\Omega_{KLM} =
-\eta_{JL}\Omega_{KIM}-\eta_{IM}\Omega_{KJL} +
\eta_{IL}\Omega_{KJM} + \eta_{JM}\Omega_{KIL} \, .
\label{master-cso}
\end{equation}
Contracting equation (\ref{master-cso}) with $\eta^{IK}\eta^{JL}$
we obtain:
\begin{equation}
\eta^{IJ}\Omega_{IJK} = 0\, , \forall K\, . \label{cso-res}
\end{equation}
Contracting (\ref{master-cso}) with $\eta^{IK}$ and using
(\ref{cso-res}) we find:
\begin{equation}
-(p+q)\Omega_{IJK} = \Omega_{IJK} + \Omega_{KIJ} + \Omega_{JIK}\,
. \label{cso-res2}
\end{equation}
Writing out (\ref{cso-res2}) three times with the indices
cyclically permuted and adding the three expressions we find the
result:
\begin{equation}
(p+q - 3)\left( \Omega_{IJK} + \Omega_{JKI} + \Omega_{KIJ} \right)
= 0\, .
\end{equation}
Since we assumed $p+q>3$ the cyclic sum of $\Omega_{IJK}$ has to
vanish.

Using the relation $[S_{IJ},v^J]=v_I$, where no sum is taken over
the repeated index $J$ and where $v^J= \eta^{JK}v_K$, and
requiring $\Omega([S_{IJ},v^J],S_{KL}) = -
\Omega(v^J,[S_{IJ},S_{KL}])$ we obtain:
\begin{equation}
\Omega_{IJK} + \Omega_{JIK} + \Omega_{KJI} = 0 \,.
\label{cso-res3}
\end{equation}

Combining (\ref{cso-res3}) and the vanishing of the cyclic sum we
see that $\Omega_{IJK} = 0$. Hence the subspace $\mf{r}$ is
orthogonal to the whole Lie algebra with respect to $\Omega$ and
$\Omega$ is degenerate. This concludes part I.

\begin{center}
\subsection*{Part II}
\end{center}
We now give for the Lie algebras listed in the lemma on invariant
metrics on $CSO$-algebras the most general invariant metric up to
a multiplicative constant.

The Lie algebra $\mf{cso}(1,0,1)$ is abelian and hence any metric
is invariant.

For the Lie algebras $\mf{cso}(2,0,2)$ and $\mf{cso}(1,1,2)$ we
use the ordered basis $\beta=\left\{j, e_1 , e_2 , f_1 , f_2 , z
\right\}$ with the only nonzero commutation relations
\begin{equation}
 \left[ j,e_a \right] = f_a\, , \ \ \left[ j,f_a\right] = \sigma
e_a\, , \ \ \left[ f_a,f_b \right] = \sigma z\,  , \ \ \left[
e_a,e_b \right] = z\, ,
\end{equation}
where $\sigma=+1$ for $\mf{cso}(1,1,2)$ and $\sigma=-1$ for
$\mf{cso}(2,0,2)$.

In the basis $\beta$ the invariant metric can be written in matrix
form as:
\begin{equation}
\Omega = \begin{pmatrix} a & 0 & 0 & 0 & 0 & +1 \\
                         0 & 0 & 0 & 0 & -1& 0 \\
                         0 & 0 & 0 & +1 & 0 & 0 \\
                         0 & 0 & +1 & 0 & 0 & 0 \\
                         0 & -1& 0 & 0 & 0 & 0 \\
                         +1 & 0 & 0 & 0 & 0 & 0 \\ \end{pmatrix}\,
                         , \ \ a\in \Real\,,
\end{equation}
for both $\mf{cso}(1,1,2)$ and $\mf{cso}(2,0,2)$. The eigenvalues
are $-1,-1,+1,+1,\tfrac{1}{2} (a+\sqrt{a^2+4})$, $\tfrac{1}{2}
(a-\sqrt{a^2+4})$ and the signature is $+++---$.

For the Lie algebras $\mf{cso}(2,1,1)$ and $\mf{cso}(3,0,1)$ we
use the ordered basis $\beta = \left\{t_1,t_2,t_3,
v_1,v_2,v_3\right\}$ such that the commutation relations are
\begin{equation}
[t_i,t_j] = \epsilon_{ijk}\eta^{kl}t_l \, , \ \
[t_i,v_j]=\epsilon_{ijk}\eta^{kl}v_l,\, \ \ [v_i,v_j]=0\,
,\label{comcso301}
\end{equation}
where $\epsilon_{ijk}$ is the three-dimensional alternating symbol
and $\eta^{ij}$ is the diagonal metric with eigenvalues
$(+1,-1,-1)$ for $\mf{cso}(2,1,1)$ and with eigenvalues
$(+1,+1,+1)$ for $\mf{cso}(3,0,1)$.

With respect to the ordered basis $\beta$ the invariant metric is
given by
\begin{equation}
\Omega = \begin{pmatrix} a\eta & \eta \\ \eta & 0 \\
\end{pmatrix}\, ,\label{omegacso301}
\end{equation}
where each entry is a $3\times 3$-matrix. The eigenvalues are
$\lambda_{\pm} =\tfrac{1}{2}(a\pm\sqrt{a^2+4})$, both with
multiplicity three, and the signature is $---+++$.

For the Lie algebra $\mf{cso}(1,0,3)$ we use the ordered basis
$\beta= \left\{ v_1,v_2,v_3,z_1,z_2,z_3\right\}$ such that the
commutation relations are
\begin{equation}
[v_i,v_j]=\tfrac{1}{2}\epsilon_{ijk}z_k\, , \ \
[v_i,z_j]=[z_i,z_j]=0\, ,
\end{equation}
where a summation is understood for every repeated index. The
invariant metric is given in matrix form with respect to the basis
$\beta$ by:
\begin{equation}
\Omega = \begin{pmatrix} A_{3\times 3} & \mathbbm{1}_{3\times 3} \\
\mathbbm{1}_{3\times 3} & 0 \\ \end{pmatrix}\,,
\end{equation}
where $A_{3\times 3}$ is an undetermined $3\times 3$-matrix. Since
$\det \Omega= -1$ there are no null vectors. We find that if
$\mu_1,\mu_2,\mu_3$ are the eigenvalues of $A$, then $\lambda_i =
\tfrac{1}{2}\Bigl( \mu_i \pm \sqrt{\mu_{i}^{2} + 4} \Bigr)\,$
are the eigenvalues of $\Omega$. Hence the signature is $+++---$.

\vspace{1truecm}
\section{The potential and its derivatives}\label{Pot}

In reference \cite{dRPTW} we presented a scheme for analyzing the
potential of $\mc{N}=4$ supergravity for semisimple gaugings. We
wish to apply this scheme for the $CSO$-gaugings, since after the
analysis of the preceding section the only difference lies in the
numerical values of the structure constants. In this section we
review the definitions and steps of the analysis of the potential.

\subsection{The potential\label{potential}}

The analysis of an extremum of the potential can be split in first
finding an extremum with respect to the $SU(1,1)$-scalars and
subsequently investigating whether the point $Z_0$ determines an
extremum with respect to the $SO(6,6)$-scalars. We therefore write
the potential as:
\begin{equation}
   V = \sum_{i,j}\, (R^{(ij)}(\phi)\, V_{ij}(Z) + I^{(ij)}(\phi)
   \,W_{ij}(Z))\,.\label{pot}
\end{equation}
The indices $i,j,\ldots$ label the different factors in the gauge
group $G$. $R^{(ij)}$ and $I^{(ij)}$ contain the $SU(1,1)$-scalars
and depend on the gauge coupling constants and the
$SU(1,1)$-angles, $V_{ij}$ and $W_{ij}$ contain the structure
constants, depend on the matter fields, and are symmetric resp.\
antisymmetric in the indices $i,j$. The $SU(1,1)$-angle associated
with the $i$th factor is written $\alpha_i$, and the structure
constants determined by the $i$th factor are denoted
$f^{(i)}_{RS}{}^T$ and we define $f^{(i)}_{RST} =
f^{(i)}_{RS}{}^U\eta_{TU}$. The functions $V_{ij}$ and $W_{ij}$
are given by:
\begin{eqnarray}
  V_{ij} &=& \tfrac{1}{4} Z^{RU}Z^{SV}(\eta^{TW}+\tfrac{2}{3}Z^{TW})\,
          f^{(i)}{}_{RST} f^{(j)}{}_{UVW}\,,  \label{Vij}
\\
  W_{ij} &=& \tfrac{1}{36}\epsilon^{abcdef}
      Z_a{}^RZ_b{}^S Z_c{}^T Z_d{}^U Z_e{}^V Z_f{}^W\,
      f^{(i)}{}_{RST} f^{(j)}{}_{UVW}\,.\label{Wij}
\end{eqnarray}

The extremum of the potential in the $SU(1,1)$-directions has been
determined in \cite{dRWP1}. For completeness we briefly review
this analysis in appendix \ref{SU11}. The value of the potential
at the extremum with respect to the $SU(1,1)$-scalars is given by
\begin{equation}
\label{pot1}
  V_0 = \sgn{C_-}\,\sqrt{\Delta}  - T_- \,,
\end{equation}
where (see \cite{dRWP1})
\begin{eqnarray}
\label{Cmin}
  C_{-} &=& \sum_{ij} g_ig_j\cos(\alpha_i-\alpha_j)V_{ij}\,,\\
\label{Tmin}
  T_{-}   &=& \sum_{ij} a_{ij} W_{ij} \,,\\
\label{Delta}
  \Delta &=& 2\,\sum_{ij}\sum_{kl} V_{ij}V_{kl} a_{ik}a_{jl}\,,\\
  \label{defaij}
  a_{ij} &\equiv& g_ig_j \sin(\alpha_i - \alpha_j)\,.
\end{eqnarray}
The condition for this extremum to exist is that $\Delta>0$, which
implies that at least two of the $SU(1,1)$-angles must be
different.

At the point $Z_0$ the functions defined above are given by
\begin{eqnarray}
V_{ij}(Z_0) &=& \delta_{ij}\Bigl(
-\tfrac{1}{12}f^{(i)}_{ABC}f^{(i)}_{ABC} + \tfrac{1}{4}
f ^{(i)}_{ABI}f ^{(i)}_{ABI} \Bigr)\label{functiesbijZ0a}\,,\\
W_{ij}(Z_0)&=&\tfrac{1}{36}\epsilon^{ABCDEF}f^{(i)}_{ABC}f^{(j)}_{DEF}\,,\\
\Delta(Z_0)&=& 2\sum_{i,j}a_{ij}^{2}V_{ii}(Z_0)V_{jj}(Z_0)\,,\\
C_-(Z_0)&=&\sum_i g_{i}^{2}V_{ii}(Z_0)\,,\\
R^{(ij)}(Z_0)&=&
\delta_{ij}\frac{2\;\textrm{sign}\;C_-(Z_0)}{\sqrt{\Delta_0}}\sum_j
V_{jj}(Z_0)a_{ij}^{2}\,, \\
I^{(ij)}(Z_0) &=& -a_{ij}\, .\label{functiesbijZ0c}
\end{eqnarray}
With the formulae (\ref{functiesbijZ0a}-\ref{functiesbijZ0c}) it is
easy to plug in the values of the structure constants and
determine the value of the potential at $Z_0$, see section
\ref{analysis}.

\subsection{The derivatives of the potential\label{firstder}}

To determine whether the point $Z_0$ is an extremum with respect
to the $SO(6,6)$-scalars we calculate the derivatives with respect
to the parameters $P_{ab}$ introduced in section \ref{scalars}
(see also \cite{dRPTW}). We have
\begin{equation}
\frac{\partial V}{\partial P_{ab}}(Z_0) =
\sum_{i}R^{(ii)}(Z_0)f^{(i)}_{a+6,CJ}f^{(i)}_{bCJ} -
\tfrac{1}{6}\sum_{ij}a_{ij}
\epsilon^{bCDEFG}f^{(i)}_{a+6,CD}f^{(j)}_{EFG}\,.
\end{equation}
Since for $CSO$-gaugings at most two groups are possible to fit in
$SO(6,6)$ the summations over the indices $i,j$ simplify
significantly.
%and the calculations are easy to implement in a computer program
%like Mathematica.
For the point $Z_0$ to be an extremum the
$6\times 6$-matrix $\partial V/\partial P$ should vanish.

If the point $Z_0$ turns out to be an extremum with respect to
both the $SU(1,1)$-scalars and the $SO(6,6)$-scalars, we need the
second derivatives at $Z_0$ to determine whether the extremum is
stable or unstable. Schematically the second derivatives are given
by
\begin{eqnarray}
\label{DDphi}
 \frac{\partial^2 V}{ \partial\phi^2} &=&
   \sum_{ij} \frac{\partial^2 R^{(ij)}}{ \partial\phi^2} V_{ij}\,,
 \\
 \label{DDphiZ}
  \frac{\partial^2 V}{ \partial\phi\partial P} &=&
   \sum_{ij} \frac{\partial R^{(ij)}}{ \partial\phi}
             \frac{\partial V_{ij}}{ \partial P} \,,
 \\
\label{DDZZ}
  \frac{\partial^2 V}{ \partial P^2} &=&
   \sum_{ij} R^{(ij)} \frac{\partial^2 V_{ij}}{ \partial P^2}
             +I^{(ij)}\frac{\partial^2 W_{ij}}{ \partial P^2}\,.
\end{eqnarray}
The second derivatives (\ref{DDphi}) were studied in \cite{dRPTW}.
The sign of (\ref{DDphi}) depends on the sign of $C_-$. For
positive (negative) $C_-$ the extremum in the $SU(1,1)$-scalars is
a minimum (maximum). The mixed second derivatives vanish if either
the derivatives with respect to the $SU(1,1)$-scalars $\phi$ or
with respect to the matter scalars vanishes.

Hence if $C_- >0$ at $Z_0$ we need to check the eigenvalues of the
matrix of second derivatives (\ref{DDZZ}). With the formulas of
reference \cite{dRPTW} it is a matter of algebra to obtain:
\begin{equation}
\begin{split}
\frac{\partial^2 V_{ij}}{\partial P_{ab}\partial P_{cd} }(Z_0) =&
0\,,
\quad i\neq j\,,\\
\frac{\partial^2 V_{ii}}{\partial P_{ab}\partial P_{cd}} (Z_0) =&
\delta_{ac}f^{(i)}_{bGJ}f^{(i)}_{dGJ} +
\delta_{bd}f^{(i)}_{a+6,GJ}f^{(i)}_{cGJ}
-\tfrac{1}{2}\delta_{bc}f^{(i)}_{a+6,GJ}f^{(i)}_{dGJ}\\&-\tfrac{1}{2}\delta_{ad}f^{(i)}_{c+6,GJ}f^{(i)}_{bGJ}
+ f^{(i)}_{a+6,c+6,R}f^{(i)}_{b,d,J} +
f^{(i)}_{b,c+6,R}f^{(i)}_{a+6,d,R}\,,\\
\frac{\partial^2 W_{ij} }{\partial P_{ab}P_{cd} } (Z_0) =&
\tfrac{1}{24}
\epsilon^{bBCDEF}\left(\delta_{ac}f^{(i)}_{dBC}f^{(j)}_{DEF}
-\delta_{ad}f^{(i)}_{c+6,BC}f^{(j)}_{DEF}\right)\\
&+\tfrac{1}{24}\epsilon^{dBCDEF}\left(\delta_{ac}f^{(i)}_{bBC}f^{(j)}_{DEF}
- \delta_{bc}f^{(i)}_{a+6,BC}f^{(j)}_{DEF} \right)\\
&+ \tfrac{1}{12} \epsilon^{bdCDEF}\left(
2f^{(i)}_{a+6,c+6,C}f^{(j)}_{DEF} + 3
f^{(i)}_{a+6,CD}f^{(j)}_{c+6,EF}\right)\\&  - (i\leftrightarrow
j)\,.
\end{split}
\end{equation}
The stability is then determined by the eigenvalues of the
$36\times 36$-matrix given by
\begin{equation}
\sum_{i}R^{(ii)}(Z_0)\frac{\partial^2 V_{ii}}{\partial
P_{ab}\partial P_{cd} } (Z_0) - \sum_{ij} a_{ij}\frac{\partial^2
W_{ij} }{\partial P_{ab}P_{cd} } (Z_0)\,.
\end{equation}

\section{Analysis of the potentials of CSO-gaugings\label{analysis}}

With the formulas of section \ref{Pot} and the structure constants
of the $CSO$-algebras, given in appendix \ref{structconst}, at our
disposal, we analyze the potential and the first and second
derivatives at $Z_0$ for different $CSO$-gaugings. For each gauge
group the function $V_{ii}(Z_0)$ is given in table \ref{tabelV0}.
Note that the value of $V_{ii}(Z_0)$ is the same for plus- and
minus-embeddings. Since only two $CSO$ gauge-algebras fit into
$SO(6,6)$ we have $\Delta(Z_0) = 4 a_{12}^{2}
V_{11}(Z_0)V_{22}(Z_0)$, hence $\Delta>0$ if and only if both
$V_{11}(Z_0)$ and $V_{22}(Z_0)$ are nonzero and have the same
sign. Hence in searching for gaugings that admit an extremum with
respect to the $SU(1,1)$-scalars, we can disregard the gaugings
that involve $CSO(2,0,2)$ and the gaugings of which precisely one
factor is $CSO(3,0,1)$.

\TABLE{
    %\begin{center}
        \begin{tabular}{|c|c||c|c|}
            \hline
                gauge factor & $V_{ii}(Z_0)$ & gauge factor &  $V_{ii}(Z_0)$ \\
                \hline
                 $CSO(1,0,3)_+$ & 1 & $CSO(1,0,3)_-$ & 1  \\
                 $CSO(2,0,2)_+$ & $0$ &  $CSO(2,0,2)_-$ & $0$  \\
                 $CSO(1,1,2)_+$ & 1 & $CSO(1,1,2)_-$ & 1 \\
                 $CSO(3,0,1)_+$& $-\tfrac{1}{2}(\lambda^4 +4\lambda^2 +1)$ &   $CSO(3,0,1)_-$ & $  -\tfrac{1}{2}(\lambda^4 +4\lambda^2
                 +1)
                 $\\
                 $CSO(2,1,1)_+$ & $ \tfrac{1}{2} ( 3\lambda^4 + 4\lambda^2 + 3 )
                 $&  $CSO(2,1,1)_-$ & $ \tfrac{1}{2} ( 3\lambda^4 + 4\lambda^2 + 3 )$\\
            \hline
        \end{tabular}\caption{The value of $V_{ii}$ at the point
        $Z_0$ for different gauge factors. The plus- and minus-sign
        refer to two distinct possibilities to embed the factor
        into the gauge group. The number $\lambda$ is an arbitrary
        positive number, coming from an arbitrary constant in the
        invariant metric.}\label{tabelV0}
    %\end{center}
}

For the groups $CSO(3,0,1)$ and $CSO(2,1,1)$ the structure
constants contain an undetermined positive parameter $\lambda$
that cannot be removed redefinition of the generators preserving
the constraints (\ref{etaconstraint}). This parameter is a remnant
of the invariant metric; there is in general an $m$-parameter
family of invariant metrics with $m>1$ for $\mf{cso}(p,q,r)$ with
$p+q+r=4$.

%The computations of $\Delta$, $C_-$, $V_{ii}(Z_0)$ and all
%derivatives with respect to $P$ of the potential are done with
%computer.

The gaugings for which the point $Z_0$ corresponds to an extremum
of both the $SU(1,1)$- and the $SO(6,6)$-scalars are:
$CSO(1,0,3)_- \times CSO(1,0,3)_-$, $CSO(1,0,3)_+\times
CSO(1,0,3)_-$ and $CSO(1,0,3)_+\times CSO(1,0,3)_+$. Only these
gaugings have vanishing derivative with respect to the parameters
$P_{ab}$ and $\Delta>0$. For these three gaugings the value of the
potential at the point $Z_0$ is given by $V_0 = 0$. With respect
to the $SU(1,1)$-scalars the potential is a minimum, $C_-(Z_0)
>0$, but with respect to the $SO(6,6)$-scalars the extremum is unstable;
the mass-matrix $\partial^2 V/\partial P\partial P$ has both
positive and negative eigenvalues.

\section{Cosmological scaling solutions}\label{scaling}

If a scalar potential is of the form
\begin{equation}
V(\chi, \Phi_i) = \e^{b\,\chi}\,U(\Phi_i)\,,
\end{equation}
where $\chi$ has canonical kinetic term and is independent of the
scalars $\Phi_i$, a cosmological scaling solution exists if the
function $U(\Phi_i)$ has a positive extremum with respect to the
scalars $\Phi_i$ \cite{Copeland:1997pi}. The scale factor of the
Friedmann-Robertson-Walker metric goes as $t^{{1}/{b^2}}$ for the
scaling solution. The characteristic feature of scaling solutions
is that the ratio of the kinetic energy of the scalar $\chi$ and
the potential energy of the scalar $\chi$ remains constant during
evolution. Scaling solutions appear as fixed points in autonomous
systems that describe scalar cosmologies, see
\cite{Copeland:2006wr} for a recent review and a list of
references.

In $\mc{N}=4$ supergravity the potential factorizes in a trivial
way if all $SU(1,1)$-angles are equal; in this case the function
$R^{(ij)}(r,\varphi)$ simplifies to:
\begin{equation}
R^{(ij)}(r,\varphi) = g_ig_j\frac{1+r^2 - 2r \cos\varphi}{1-r^2} =
g_ig_j\frac{|1+z|^2}{1-|z|^2}\,,
\end{equation}
where $z=-r\e^{i\varphi}$. Introducing $\tau = i (1-z)/(1+z)$,
which takes values in the complex upper half plane since $|z|<1$,
and $\sigma = \textrm{Re}\tau$ and $\e^{-\chi} = \textrm{Im}\tau$
one finds
\begin{equation}
R^{(ij)}(\chi,\sigma)  = g_i g_j \e^{\chi} \,.
\end{equation}

Hence we find for the potential at $Z_0$ in this case
\begin{equation}
V(Z_0) = -\tfrac{1}{12}\e^{\chi} \sum_{i}g_{i}^{2}\left(
f^{(i)}_{ABC}f^{(i)}_{ABC} - 3
f^{(i)}_{ABI}f^{(i)}_{ABI}\right)\,.
\end{equation}
The first derivatives with respect to $P_{ab}$ at $Z_0$ simplifies
to:
\begin{equation}
\frac{\partial V}{\partial P_{ab}}(Z_0) = \e^{\chi}
\sum_{i}g_{i}^{2} f^{(i)}_{a+6,DK}f^{(i)}_{bDK}\,.
\end{equation}
The second derivatives with respect to $P_{ab}$ at $Z_0$ become:
\begin{equation}
\begin{split}
\frac{\partial^2 V}{\partial P_{ab} \partial P_{cd}}(Z_0) = &
\e^{\chi}\sum_{i}g_{i}^{2} \Bigl( \delta_{ac}
f^{(i)}_{bCJ}f^{(i)}_{dCJ} +
\delta_{bd}f^{(i)}_{a+6,CJ}f^{(i)}_{c+6,CJ} \\
& -\tfrac{1}{2}
\delta_{bc}f^{(i)}_{a+6,CJ}f^{(i)}_{dCJ}-\tfrac{1}{2}
\delta_{ad}f^{(i)}_{c+6,CJ}f^{(i)}_{bCJ} + 2
f^{(i)}_{a+6,c+6,R}f^{(i)}_{bdR} \Bigr)
\end{split}
\end{equation}

%Again the computation is straightforward and most easily done with
%the aid of a computer.
\noindent The computations are simplified by
noting that the formulas factorize into contributions of different
factor groups. Hence to look for an extremum one only has to
investigate the contributions of different factor groups to
$\partial V/\partial P$.

We find that only $CSO(1,1,2)$ has vanishing contribution to
$\partial V/\partial P$ and hence we find that the $CSO$-gaugings
that allow for scaling solutions at $Z_0$ are $CSO(1,1,2)$ and
$CSO(1,1,2)\times CSO(1,1,2)$. Note that the structure constants
of $CSO(1,1,2)_+ $ are the same as of $CSO(1,1,2)_-$. For the
gauging $CSO(1,1,2)\times CSO(1,1,2)$ the eigenvalues of
$\partial^2 V/\partial P^2$ are found to be all positive. The
potential at $Z_0$ is given by:
\begin{equation}
V(\chi,Z_0) = (g_{1}^{2} + g_{2}^{2})\,\e^\chi\,.
\end{equation}
Hence the gauging $CSO(1,1,2)\times CSO(1,1,2)$ admits a stable
scaling solution. The same is then true for the gauging $CSO(1,1,2)$,
since this is a truncation of the gauging $CSO(1,1,2)\times CSO(1,1,2)$
obtained by putting $g_2=0$.

\section{Conclusions}

The conclusions of this paper can be split in three parts.

The first conclusion concerns the gaugings in matter-coupled
$\mc{N}=4$ supergravity with $CSO$-groups. In the formulation of
$\mc{N}=4$ supergravity of \cite{MdRPW1} the only possible
$CSO$-gaugings require that the Lie algebra $\mf{cso}(p,q,r)$
admits an invariant metric. The only Lie algebras
$\mf{cso}(p,q,r)$ with $r>0$ that admit an invariant metric are
those with $p+q+r = 2,4$. If $p+q+r=2$ the Lie algebra
$\mf{cso}(p,q,r)$ is abelian and hence we considered only
$p+q+r=4$.

The second conclusion is that the $CSO$-gaugings that we
considered showed no stable minimum with respect to all $36+2$
scalars at the point $Z_0$.
This analysis concerns the case of $\mc{N}=4$ supergravity with
six vectormultiplets, and is therefore not completely general.
Also the formalism used in the present paper and in \cite{dRPTW}
has recently been generalized \cite{Schon:2006kz}. Going beyond the
present paper as proposed in \cite{Schon:2006kz} involves
solving a system of constraints involving  parameters
$\left\{ \xi_{\alpha M},\ f_{\alpha KLM} \right\}$.
It is an interesting and important challenge to solve these
equations for $\xi_{\alpha M}\ne 0$, and to perform a general
analysis of scalar potentials in gauged $\mc{N}=4$ supergravity.

The third conclusion is that a stable scaling solution exists at
$Z_0$ in $\mc{N}=4$ gauged supergravity with gauge group
$CSO(1,1,2)$, or any power of $CSO(1,1,2)$. The scaling solution
is characterized by a scale factor, which grows linearly in time
and the effective potential contains one scalar $\chi$;
\begin{equation}
V_{eff}(\chi) = (g_{1}^{2}+g_{2}^{2}+\ldots )\,\e^\chi\,.
\end{equation}
The numbers $g_i$ are the coupling constants for each factor of
$CSO(1,1,2)$. Also this analysis is not exhaustive. For example,
there might be scalars in the $SO(6,6)$-sector that factorize out
of the potential such as to combine with the $SU(1,1)$-scalar an
overall exponential factor. It will be interesting to study
the cosmological models resulting from these scaling solutions
in more detail.
%But also the formulation of
%\cite{Schon:2006kz} can shed new light on the subject.

\acknowledgments

\bigskip

The work of DBW is part of the research programme of the
``Stichting voor Fundamenteel Onderzoek van de Materie'' (FOM). SP
thanks  the Centre for Theoretical Physics in Groningen for their
hospitality. MdR and DBW are supported by the European
Commission FP6 program MRTN-CT-2004-005104 in which MdR and DBW
are associated to Utrecht University.

\appendix

\section{$SU(1,1)$ scalars and angles\label{SU11}}
When we parameterize the coset $SU(1,1)/U(1)$ as in
eqn.(\ref{paramSU}), the scalars $r$ and $\varphi$ appear in the
potential (\ref{pot}) through
\begin{eqnarray}
  R^{(ij)} &=& \frac{g_i g_j}{ 2}  (\Phi^*_i\Phi_j + \Phi^*_j\Phi_i)\nonumber\\
           &=& g_ig_j\left(\cos(\alpha_i-\alpha_j)\frac{1+r^2}{ 1-r^2}
                - \frac{2r}{ 1-r^2}\cos(\alpha_i+\alpha_j+\varphi)
                \right)\,,\\
  I^{(ij)} &=& \frac{g_ig_j }{ 2i} (\Phi^*_i\Phi_j - \Phi^*_j\Phi_i)= -g_ig_j\sin(\alpha_i-\alpha_j)\,.
\end{eqnarray}
Introducing
\begin{eqnarray}
\label{CS}
  C_{\pm} &=& \sum_{ij} g_ig_j\cos(\alpha_i\pm\alpha_j)V_{ij}\,,\quad
  S_{+} = \sum_{ij} g_ig_j\sin(\alpha_i+\alpha_j)V_{ij} \,,\\
  T_{-}   &=& \sum_{ij} g_ig_j\sin(\alpha_i-\alpha_j)W_{ij} \,,
\end{eqnarray}
we rewrite the potential as
\begin{equation}
\label{potrphi}
   V = C_-\,\frac{1+r^2}{ 1-r^2} - \frac{2r}{ 1-r^2}\,
     \big(C_+\cos\varphi - S_+\sin\varphi\big) - T_-\,.
\end{equation}
This extremum in $r$ and $\varphi$ takes on the form
\begin{eqnarray}
  \cos\varphi_0 &=& \frac{s_1  C_+}{ \sqrt{C_+^2 + S_+^2}}\,,\quad
  \sin\varphi_0 = -\frac{s_1 S_+}{ \sqrt{C_+^2 + S_+^2}}\,,
  \nonumber\\
    r_0 &=& \frac{1}{ \sqrt{C_+^2 + S_+^2}}
    \left( s_1 C_- + s_2 \sqrt{\Delta}\right)\,,\qquad
    \Delta\equiv C_-^2-C_+^2-S_+^2\,,
\end{eqnarray}
where $s_1$ and $s_2$ are signs. These are determined by requiring
$0\leq r_0<1$, this gives $s_1=\sgn C_-$ and $s_2=-1$.
Substitution of $r_0$ and $\varphi_0$ in $V$ leads to
eqn.(\ref{pot1}).

In the case that all $SU(1,1)$ angles $\alpha_i$ vanish,
$S_+=T_-=0$ and $C_-=C_+$, and one finds $r_0=1$ and $\Delta=0$.
This is a singular point of the parametrization, which we will
exclude. This case corresponds to the Freedman-Schwarz potential
\cite{FS}, which has no minimum.

For the kinetic term and mass-matrix of the $SU(1,1)$-scalars we
introduce:
\begin{eqnarray}
x &=& \frac{2}{ (1-r_0)^2}(r\cos\varphi - r_0\cos\varphi_0)\,,
 \nonumber\\
y &=& \frac{2}{ (1-r_0)^2}(r\sin\varphi - r_0\sin\varphi_0)\,.
\end{eqnarray}
In these variables we find
\begin{eqnarray}
\label{actionphi}
   {\cal L}(x,y) &=&
   - \tfrac{1}{2} \left( \frac{1-r_0^2}{ 1-r^2}\right)^2
    \big( (\partial x)^2 + (\partial y)^2)
   - V_0\nonumber\\
   &&-\tfrac{1}{2}\, \sgn{C_-}\,\sqrt{\Delta}\,( x^{2}
    + y^2 ) +\ldots\,,
\end{eqnarray}
where the ellipsis indicate terms of higher order in $x$ and $y$.

\section{Structure constants}\label{structconst}

In this appendix we give the structure constants of the
$\mf{cso}(p,q,r)$ Lie algebras with $p+q+r=4$ in a basis such that
the constraint (\ref{etaconstraint}) is satisfied. The Lie
algebras $\mf{cso}(p,q,r)$ with $p+q+r=4$ have dimension six and
the invariant metric has signature $+++---$. A gauge algebra
consists of two Lie algebras $\mf{cso}(p,q,r)$ with $p+q+r=4$, and
the first Lie algebra can be embedded into the subspace spanned by
the generators $T_1,T_2,T_3,T_7,T_8,T_9$ and the second can
embedded into the subspace spanned by the generators
$T_4,T_5,T_6,T_{10},T_{11},T_{12}$.

We give the structure constants of every $\mf{cso}(p,q,r)$ with
$p+q+r=4$ as embedded in the subspace spanned by the generators
$T_1,T_2,T_3,T_7,T_8,T_9$ since the other embedding can be
obtained from the latter by the following permutation of the
indices: $\sigma=(14)(25)(36)(7\; 10)(8\; 11)(9\; 12)\in S_{12}$.
In fact we also only give the structure constants of the
plus-embedding, the minus-embedding (with the generators lying in
the same subspace) can be obtained by applying the following
permutation of the indices: $\tau = (17)(28)(39)(4\; 10)(5\;
11)(6\; 12) \in S_{12}$. Consistency requires $\sigma\tau =
\tau\sigma$, which is easily seen to be satisfied.

With these preliminaries the structure constants of
$\mf{cso}(2,0,2)$, $\mf{cso}(1,1,2)$ and $\mf{cso}(1,0,3)$ are
given as in table \ref{tabelstruct}. To be economic in writing we
only present the nonzero structure constants $f_{RS}{}^T$ for
which $R<S$.

\TABLE[ht]{
    %\begin{center}
        \begin{tabular}[ht]{|c|c|c|c|c|}
            \hline
            \multicolumn{5}{|c|}{$\mathbf{\mf{cso}(2,0,2)}$} \\
            \hline
            $f_{12}{}^3 = 1$ & $f_{13}{}^2 =-1$ & $f_{18}{}^9=1$
            & $f_{19}{}^8 = -1 $ & $ f_{23}{}^1=1$\\
            $f_{23}{}^7=-1$ & $f_{27}{}^3=-1$ & $f_{37}{}^2=1$ & $
            f_{78}{}^9 = 1$ & $f_{79}{}^8 = -1$ \\
            $f_{89}{}^1 = -1$ & $ f_{56}{}^7 = 1$ &&&\\
            \hline
            \hline
            \multicolumn{5}{|c|}{$\mathbf{\mf{cso}(1,1,2)}$} \\
            \hline
            $f_{12}{}^8 = -1$ & $f_{13}{}^9 =1$ & $f_{18}{}^2=-1$
            & $f_{19}{}^3 = 1 $ & $ f_{27}{}^8=1$\\
            $f_{28}{}^1=1$ & $f_{28}{}^7=-1$ & $f_{37}{}^9=-1$ & $
            f_{39}{}^1 = -1$ & $f_{39}{}^7 = 1$ \\
            $f_{78}{}^2 = -1$ & $ f_{79}{}^3 = 1$ &&&\\
            \hline
            \hline
            \multicolumn{5}{|c|}{$\mathbf{\mf{cso}(1,0,3)}$} \\
            \hline
            $f_{12}{}^3 = 1$ & $f_{12}{}^9 =-1$ & $f_{18}{}^9=-1$
            & $f_{13}{}^2 = -1 $ & $ f_{13}{}^8=1$\\
            $f_{18}{}^3=1$ & $f_{19}{}^2=-1$ & $f_{19}{}^8=1$ & $
            f_{23}{}^1 = 1$ & $f_{23}{}^7 = -1$ \\
            $f_{27}{}^3 = -1$ & $ f_{27}{}^9 = 1$ & $f_{29}{}^1=1 $& $f_{29}{}^7 = -1$ &  $f_{37}{}^2 =1$ \\
            $ f_{27}{}^8 =-1$ & $ f_{38}{}^1 =-1 $ & $ f_{38}{}^7 =1
            $& $ f_{78}{}^3 = 1$ & $f_{78}{}^9 =-1  $ \\
            $f_{79}{}^2 =-1  $ & $f_{79}{}^5 = 1$  & $f_{89}{}^1 =1 $  & $f_{89}{}^7 =-1 $ & \\
            \hline
            \hline
            \multicolumn{5}{|c|}{$\mathbf{\mf{cso}(2,1,1)}$} \\
            \hline
            $f_{12}{}^3 = -\lambda^2$ & $f_{12}{}^9 =-1$ & $f_{13}{}^2=\lambda^2$
            & $f_{13}{}^8 = 1 $ & $ f_{18}{}^3=1$\\
            $f_{18}{}^9=-(\lambda^2 +2) $ & $f_{19}{}^2=-1$ & $f_{19}{}^8=(\lambda^2+2)$ & $
            f_{23}{}^1 = -\lambda^2$ & $f_{23}{}^7 = 2\lambda^2+1$ \\
            $f_{27}{}^3 = 2\lambda^2+1$ & $ f_{27}{}^9 = -\lambda^2$ & $f_{29}{}^1=1 $& $f_{29}{}^7 = \lambda^2$ &  $f_{37}{}^2 =-(2\lambda^2+1)$ \\
            $ f_{37}{}^8 =\lambda^2$ & $ f_{38}{}^1 =-1 $ & $ f_{38}{}^7 =\lambda^2
            $& $ f_{78}{}^3 = -\lambda^2$ & $f_{78}{}^9 =-1  $ \\
            $f_{79}{}^2 =\lambda^2  $ & $f_{79}{}^5 = 1$  & $f_{89}{}^1 =\lambda^2+2 $  & $f_{89}{}^7 =-1 $ & \\
            \hline
        \end{tabular}\caption{Structure constants of some relevant
        $\mf{cso}$-algebras.}\label{tabelstruct}
   % \end{center}
}

The number $\lambda$ is related to the undetermined constant $a$
in the invariant metric of $\mf{cs}(2,1,1)$ and $\mf{cso}(3,0,1)$
by $2\lambda = a+\sqrt{a^2 + 4}$. Since the function $x\mapsto
x+\sqrt{x^2 + 4}$ is one-to-one from $\Real$ to the set of
positive real numbers, the number $\lambda$ can be considered an
arbitrary positive real number.

The totally antisymmetric tensors $f_{ABC} = f_{AB}{}^D \eta_{DC}$
of $\mf{cso}(3,0,1)$ are more easily displayed in tensor form:
\begin{equation}
 \begin{split}
    f_{ABC}  = -(\lambda^2 +2) \epsilon_{ABC} \,,  \ \ \
    &f_{ABI}  = -\epsilon_{AB(I-6)}\,, \\
    f_{AIJ}  = \lambda^2 \epsilon_{A(I-6)(J-6)}\,, \ \ \
    &f_{IJK}  = (2\lambda^2 +1)\epsilon_{(I-6)(J-6)(K-6)}\,,
 \end{split}
\end{equation}
where $\epsilon_{abc} = +1 (-1)$ if $(abc)$ is an even (odd)
permutation of $(123)$, and otherwise it is zero.

\end{document}